\documentclass[aps,pra,amsmath,twocolumn,amssymb,floatfix,showpacs,superscriptaddress,nofootinbib,noeprint,nourl]{revtex4-1}
\usepackage{tikz}
\usepackage{graphicx}%
\usepackage{color} 
\usepackage{xcolor}
\usepackage{transparent}
\usepackage{psfrag}
\usepackage{dcolumn}
\usepackage{amsmath,wasysym}
\usepackage{bm} 
\usepackage{amssymb}
\usepackage{latexsym}
\usepackage{booktabs}
\usepackage{latexsym, bbold,bm}
\usepackage[colorlinks=true,linktoc=page,linkcolor=blue,citecolor=blue,urlcolor=blue]{hyperref}

\begin{document}


\definecolor{cream}{RGB}{222,217,201}
\definecolor{forestgreen}{rgb}{0.13, 0.55, 0.13}
\definecolor{brilliantrose}{rgb}{1.0, 0.33, 0.64}
\definecolor{darkmagenta}{RGB}{39, 0, 139}

\def\bea{\begin{eqnarray}}
\def\eea{\end{eqnarray}}
\def\beq{\begin{equation}}
\def\eeq{\end{equation}}
\def\f{\frac}
\def\k{\kappa}
\def\e{\epsilon}
\def\ve{\varepsilon}
\def\be{\beta}
\def\D{\Delta}
\def\h{\theta}
\def\t{\tau}
\def\a{\alpha}

\def\cDa{{\cal D}[X]}
\def\cD{{\cal D}[x]}
\def\cL{{\cal L}}
\def\cLo{{\cal L}_0}
\def\cLa{{\cal L}_1}
\def\rv{{\bf r}}
\def\tv{\hat t}
\def\on{{\omega_{\rm a}}}
\def\od{{\omega_{\rm d}}}
\def\off{{\omega_{\rm off}}}
\def\fv{{\bf{f}}}
\def\fm{\bf{f}_m}
\def\zh{\hat{z}}
\def\yh{\hat{y}}
\def\xh{\hat{x}}
\def\km{k_{m}}
\def\nh{\hat{n}}

\def\Re{{\rm Re}}
\def\sj{\sum_{j=1}^2}
\def\rk{\rho^{ (k) }}
\def\rek{\rho^{ (1) }}
\def\cek{C^{ (1) }}
\def\rz{\rho^{ (0) }}
\def\rt{\rho^{ (2) }}
\def\rtb{\bar \rho^{ (2) }}
\def\trk{\tilde\rho^{ (k) }}
\def\trek{\tilde\rho^{ (1) }}
\def\trz{\tilde\rho^{ (0) }}
\def\trt{\tilde\rho^{ (2) }}
\def\r{\rho}
\def\tD{\tilde {D}}

\def\s{\sigma}
\def\kb{k_B}
\def\bF{\bar{\cal F}}
\def\F{{\cal F}}
\def\la{\langle}
\def\ra{\rangle}
\def\nn{\nonumber}
\def\up{\uparrow}
\def\dn{\downarrow}
\def\S{\Sigma}
\def\dg{\dagger}
\def\d{\delta}
\def\p{\partial}
\def\l{\lambda}
\def\L{\Lambda}
\def\G{\Gamma}
\def\o{\Omega}
\def\w{\omega}
\def\g{\gamma}
\def\E{{\mathcal E}}

\def\O{\Omega}

\def\vv{ {\bf v}}
\def\jv{ {\bf j}}
\def\jr{ {\bf j}_r}
\def\jd{ {\bf j}_d}
\def\jdd{ { j}_d}
\def\noi{\noindent}
\def\a{\alpha}
\def\d{\delta}
\def\p{\partial} 

\def\la{\langle}
\def\ra{\rangle}
\def\e{\epsilon}
\def\n{\eta}
\def\g{\gamma}
\def\break#1{\pagebreak \vspace*{#1}}
\def\hf{\frac{1}{2}}
\def\rcurs{r_{ij}}

\def\bv{ {\bf b}}
\def\uv{ {\bf u}}
\def\rv{ {\bf r}}
\def\cf{{\mathcal F}}



\title{Cooperation and competition in the collective drive by motor proteins: Mean active force, fluctuations, and self-load}
\author{Chitrak Karan}
\email{chitrak.k@iopb.res.in}
\affiliation{Institute of Physics, Sachivalaya Marg, Sainik School, Bhubaneswar 751005, India}
\affiliation{Homi Bhaba National Institute, Training School Complex, Anushakti Nagar, Mumbai 400094, India}

\author{Debasish Chaudhuri}
\email{debc@iopb.res.in}
\affiliation{Institute of Physics, Sachivalaya Marg, Sainik School, Bhubaneswar 751005, India}
\affiliation{Homi Bhaba National Institute, Training School Complex, Anushakti Nagar, Mumbai 400094, India}

\date{\today}

\begin{abstract}
We consider the dynamics of a bio-filament under the collective drive of motor proteins. They are attached irreversibly to a substrate and undergo stochastic attachment-detachment with the filament to produce a directed force on it. We establish the dependence of the mean directed force and force correlations on the parameters describing the individual motor proteins using analytical theory and direct numerical simulations. The effective Langevin description for the filament motion gives mean-squared displacement, asymptotic diffusion constant, and mobility leading to an effective temperature. Finally, we show how competition between motor protein extensions generates a self-load, describable in terms of the effective temperature, affecting the filament motion. 
\end{abstract}

\maketitle

\section{Introduction}
\label{sec_intro}

Motor proteins (MP) are an integral part of the cytoskeleton in eukaryotic cells~\cite{alberts2018essential, howard2001mechanics, Mugnai2020}. They are involved in a wide span of functions in subcellular processes, e.g., intracellular transport of cargo, cytoskeletal dynamics, stress generation, and cell locomotion. They hydrolyze ATP to undergo attachment-detachment and perform directed motion along conjugate filaments in the attached state~\cite{Julicher1997, Vale2003, Kolomeisky2007a, Chowdhury2013a, Brown2020}. For example, kinesin and dynein families of MPs move along microtubules, and the myosin family of MPs can move along filamentous actins. Their motion is load-dependent~\cite{Block2003, Carter2005} and the maximum velocity they can attain is subject to the available ATP concentration~\cite{Schnitzer2000}.  The local dissipation of chemical potential by ATP hydrolysis drives MPs out of equilibrium. Their direction of motion is determined by the local front-back asymmetry of conjugate filaments they can walk on. Generating non-equilibrium drive at the smallest scales, MPs constitute a class of active matter~\cite{Julicher2018, Needleman2017, Marchetti2013} in which the time-reversal symmetry and equilibrium fluctuation-dissipation relations are broken.

In living cells, MPs work together to transport various cargo, including organelles~\cite{Gross2002,  Hill2004, hancock2008intracellular, Shtridelman2008, Holzbaur2010}.
From a few to hundreds of MPs can participate in such transport~\cite{Leopold1992, Svoboda1994, Soppina2009, Rai2013b, Derr2012, Furuta2013}. 
Theoretical studies of multiple MP-driven cargo dynamics use either equal load sharing approximation or detailed numerical simulations of a finite number of MPs~\cite{Klumpp2005, Klumpp2008, Kunwar2010, Campas2005, Leduc2010b, Bhat2016, Bhat2017, Leighton2022}. 
The coupling between MPs can arise from a direct mechanical linkage as in myosin filaments~\citep{Linari2015},  molecular crowding effects~\cite{Miedema2017, Leduc2012}, or binding to cargo, the possible impacts of which have not yet been completely understood. Elastically coupled MPs show strain-induced unbinding and stalling~\cite{Berger2012,Berger2013,Berger2015}. For weak coupling, effective unbinding rate and average cargo velocity return to the non-interacting limit of single motor behavior.
In addition to performing intracellular transport, MPs can produce local active stress by sliding filaments against membrane or other filaments~\cite{Fletcher2010, Laan2012, Julicher2018, Shee2021a}. Thereby MPs promote the organization and dynamics of the mitotic spindle and positioning of microtubule asters~\cite{Grill2005, Laan2012, Ghosh2017, Pavin2021}.

Important insights into the working of MPs have been gained from {\em in vitro} gliding assay experiments~\cite{Kron1986, Howard1989, Vale1994, Bourdieu1995, Lam2014, Korten2018, Reuther2021}. 
In them, the MP tails are attached irreversibly to a substrate. The head domains of MPs can attach to conjugate filaments and, while walking on them, drive the filaments in the opposite direction. This generates an active motion of filaments. 
The motion of actin filaments driven by myosin bed showed two intriguing properties~\cite{Bourdieu1995, Lam2014}. The speed of the filament increases to saturate as the density of myosins increases. Moreover, pinned filaments show spiraling motion at high MP density~\cite{Bourdieu1995, Lam2014}.
Cooperation and competition between MPs in driving cargo generate rich dynamics~\cite{Leduc2010a, Uccar2017, Uccar2019, Scharrel2014, Grover2016, Braun2017,  Reuther2021a, DSouza2022}.
In a large assembly of F-actins or microtubules driven by such an MP bed of conjugate MPs, intriguing collective motion, including gliding, swirling, and spiral formation, was observed~\cite{AMOS1991, Sekimoto1995, Sumino2012, Schaller2010}.   
The motion of MP-driven semiflexible filaments led to several remarkable properties, including dynamical transitions between spiral and open chain conformations~\cite{Shee2021, Gupta2019a, Chaudhuri2016j}. Other active polymer models with the tangential drive led to similar behaviors~\cite{Jiang2014b, Isele-Holder2015d, Man2019, Winkler2020}. However, in the absence of direct mapping, it remains unclear what kind of MP activity can lead to which dynamical property of filaments.

In this paper, we consider the dynamics of a rigid filament driven by a gliding assay of conjugate MPs.  We develop an active bath description of the filament motion identifying and characterizing the mean force and force fluctuations due to the MP assay. We find that a lack of synchrony in the MP extension generates a self-load, reducing the efficiency with which many MPs can drive the filament together.  For this, we utilize direct numerical simulations of the stochastic dynamics of MPs and the filament and formulate an approximate analytic theory using a mean-field approach. Our work builds on the recent extensions of thermodynamic concepts to active matter~\cite{Solon2015b, Mandal2019, Petrelli2020, Loi2010}, and descriptions of tracer dynamics in active particle bath~\cite{Reichert2021, Rizkallah2022}.  Our main achievements in this paper are the following: (i)~We obtain approximate analytic expressions for the mean force and force correlation. (ii)~We derive an approximate expression for the self-load utilizing an effective temperature. Our first result directly connects the active forces felt by the filament to MP properties. It can be utilized in future active polymer modeling for the many-body dynamics of MP-driven biological filaments. Direct comparisons between our numerical results to analytic expressions show good quantitative agreement in several cases and qualitative agreement in others.

In section~\ref{sec_model} we present the detailed model. Results of numerical simulations, derivations of approximate analytic expressions for mean force and force correlations, and fluctuation-response are presented in sections~\ref{sub_sec_MFT} and~\ref{sub_sec_FDT}. The appearance of self-load is discussed in section~\ref{sub_sec_self_load}. Finally, we conclude in section~\ref{sec_discussion} by summarizing the main results.
\section{Model}
\label{sec_model}

\begin{figure}[t] 
\includegraphics[width=8.6cm]{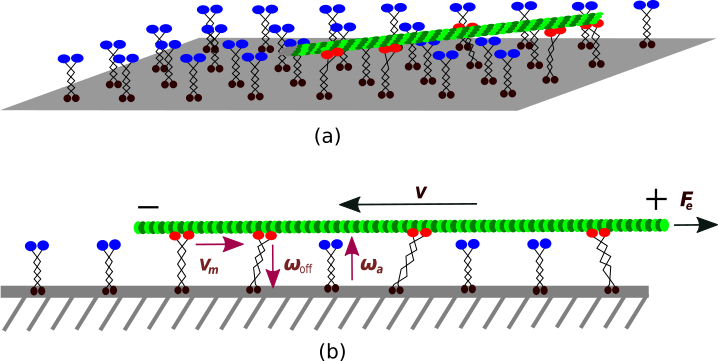}
\caption{ (color online) (a)~Schematic diagram of a gliding assay of MPs driving a conjugate filament. MP heads attached to the filament are denoted by red, and that detached from the filament are shown in blue. (b)~Side view: An attached kinesin (red) walks along the filament towards the plus end with velocity $v_m$, pulling the filament in the opposite direction. $F_e$ denotes a possible external force. The filament velocity is $v$. The MPs undergo attachment-detachment with rates $\on$ and $\off$, respectively.}
\label{fig_schematic}
\end{figure}

We consider a gliding assay (Fig.\ref{fig_schematic}) in which the tails of  MPs are attached irreversibly to a substrate. MP stems are assumed to be active extensile springs of stiffness $k_m$. The MP heads can bind to the filament with a constant rate $\on$ in a diffusion-limited manner. The attached MP head can extend along the filament in a directed fashion. This requires an active extension of MPs consuming energy from ATP hydrolysis. The extension velocity of $i$-th MP, $v_m^i$, depends on the load force $f_l^i=k_m y^i$ due to its extension $y^i$ . We adopt a piece-wise linear form of the force-dependent velocity of MPs~\cite{Leduc2010a, Carter2005},
\bea
v_m^i(f_l^i) =
\begin{cases}
        v_0 & \text{for } f_l^i \leq 0 \\
        v_0 \left( 1 - \f{f_l^i}{f_s} \right) & \text{for } 0 < f_l^i < f_b, f_b > f_s \\
        -v_{back} & \text{for } f_l^i > f_b
\end{cases}
\label{eq_piece_wise_linear_velocity}
\eea
where $f_s$ is the stall force, and $v_0$ denotes the intrinsic MP velocity. For a load force beyond the stall force, $f_l \ge f_b > f_s$, the velocity saturates to an extremely small negative value, $-v_{back}$~\cite{Leduc2010a,Carter2005}. Supportive loads do not affect the intrinsic MP motion. Assuming slip bonds, MPs can detach from the filament with a load-dependent rate, $\w_{\rm off} = \w_d \exp(|f_l^i|/f_d )$. These attachment-detachment kinetics break detailed balance. An attached motor moves along the filament stochastically with a rate $v_m/\s$, where $\s$ is the step size of the motion.

The mechanical force balance determines the over-damped dynamics of the filament position $x$,
\bea
\g_f \dot x = F_m + F_e,
\label{eq_xdot_force_balance}
\eea
where the left-hand side corresponds to the friction force, characterized by $\g_f$ and the associated motion of the filament, $\dot x$, relative to the substrate. 
The $n_a$ number of attached MPs exert a total force $F_m = - \sum_{i=1}^{n_a} f_l^i$. Here, $F_e$ denotes any external force acting on the filament. The filament motion can drag the attached MPs along with it.  Thus the extension of the $i$-th MP evolves as
\bea
\dot y^i = v_m^i(f_l^i) + \dot x.
\label{eq_ydot_ith}
\eea
\begin{table}[htbp]
\begin{center}
\caption{Parameters: The values correspond to kinesin-1 MP at ATP concentrations of 2\,mM. $\g_f$ denotes the longitudinal drag coefficient of a microtubule of length $\sim 10\,\mu$m~\cite{Lansky2015, Ghosh2017}.}
\begin{tabular}{c c c c}
\hline\hline
Definition & Parameters & Values \\
\hline
active velocity  & $v_0$ & $0.4~\mu$m/s~\cite{Schnitzer2000,Block2003}\\
stall force  &  $f_s$  &  $7.5$ pN~\cite{Carter2005,Block2003}\\
back velocity  &  $v_{back}$  &  $0.02~\mu$m/s~\cite{Carter2005}\\
detachment force  &  $f_d$ &  2.4 pN~\cite{Ghosh2017}\\
attachment rate  &  $\on$  &  5 /s~\cite{Schnitzer2000,Scharrel2014}\\
detachment rate &  $\w_d$  &  1 /s~\cite{Block2003}\\
motor stiffness  &  $k_m$   &  300 pN/$\mu$m~\cite{Kawaguchi2001}\\
MT viscous friction  &  $\g_f$  &  3.75 pN-s/$\mu$m~\cite{Lansky2015}\\
motor step-size  &  $\s$  &  0.008 $\mu$m~\cite{Svoboda1993}\\
\hline\hline
\end{tabular}
\label{tab_parameter_values}
\end{center}
\end{table}

In the simulation, we discretize the one-dimensional filament into beads separated by a length $\s$, chosen to be the same as the MP step-size. Such a discretization is considered to incorporate a capture radius $r_c=\s/2$ for the heads of detached MPs to attach to a nearby filament segment with a rate $\on$. The attached head of $i$-th MP moves unidirectionally in a stochastic manner with hopping rate $v_m^i/\s$ and by a step-size $\s$. The resultant extension of the MP produces an active force on the filament. All such forces add up to external force to displace the filament position. MPs detach from the filament with a rate $\off$ that depends on the extension $y^i$ as outlined above. We consider the filament to have a length $L=10^3 \s$. The separation between the consecutive positions to which MP tails are irreversibly attached is $L/N$. We vary $N$, keeping $L$ constant to change the MP density.

To express the dynamical quantities in a dimensionless form we use time scale $\w_d^{-1}$, length scale $v_0 \w_d^{-1}$ and force scale $\g_f v_0$. We get $\tilde{t} = t\w_d$, $\tilde{x} = x\w_d/v_0$, $\tilde{v} = v/v_0$, $\tilde{f} = \bar{f}/\g_f v_0$, $\tilde{k}_m = k_m /\g_f \w_d$. We perform Euler integration of Eq.(\ref{eq_xdot_force_balance}). The attachment-detachment and displacement of $i$-th MP position are updated using the Monte-Carlo method governed by rates $\on$, $\off$ and $v_m^i/\s$ respectively. In updating the actual $y^i$s, filament displacement is also added. We perform numerical simulations using experimentally measured parameter values for kinesins and microtubules shown in Table-\ref{tab_parameter_values}. Unless specified otherwise, the numerical results use the values listed in the table.
\section{Results}
\label{sec_results}

\subsection{Active Langevin motion}
\label{sub_sec_MFT}
\begin{figure}[t] 
\includegraphics[width=8.6cm]{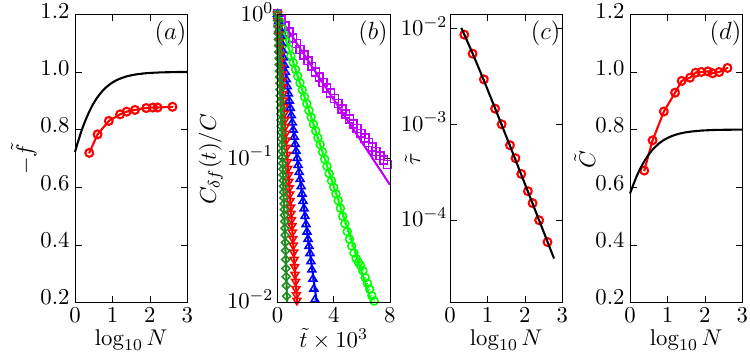}
\caption{(color online) Properties of mean force and force fluctuations. We plot $\tilde{f} = \bar{f}/(\g_f v_0)$, $\tilde{\tau} = \tau \w_d$, $\tilde{C} = C/(\g_f v_0)^2$. 
$(a)$~The points show simulation results for $-\tilde{f}$ as a function of $N$. The black solid line plots Eq.(\ref{eq_fbar}) with $\o=0.52$. 
$(b)$~Two-time correlation of force fluctuations $C_{\d f}(t)$ obtained from numerical simulations using $N=8$\,({\color{darkmagenta} $\boxdot$}), $16$\,({\color{green} $\odot$}), $40$\,({\color{blue} $\triangle$}), $80$\,({\color{red} \large{$\triangledown$}}),  $160$\, ({\color{forestgreen} \LARGE{$\diamond$}}).  The solid lines show exponential fits for correlation times $\t$.  
$(c)$~Points denote the numerically obtained correlation times and the black solid line is the analytical plot of $\t=\t_{el}$ using Eq.(\ref{eq_tau_el}).
$(d)$~Points denote  $\tilde{C}$ obtained numerically, and the black solid line plots Eq.(\ref{eq_C}).
For the analytical plots shown above we substitute $\la n_a \ra$ by $\bar n_a$ with $\o=0.52$.
}
\label{fig_results1}
\end{figure}

From numerical simulations, we find that the dynamics of the filament driven by MPs can be expressed in terms of the following Langevin equation
\bea
\g_f \dot x = f(t) = \bar f + \d f(t),
\label{eq_xdot_effective}
\eea
where the mean force $\la f(t) \ra = \bar f$. The stochastic fluctuation $\d f = f(t)-\bar f$ has the mean $\la \d f(t) \ra = 0$, and shows an exponential correlation $C_{\d f(t)}=\la \d f(t) \d f(t') \ra = C e^{-|t-t'|/\t}$ with $C = \la \d f^2(0) \ra$ (see Fig.\ref{fig_results1}). Such a colored noise arises from an underlying Ornstein-Uhlenbeck process with relaxation time $\t$~\cite{Gardiner_book}. In the rest of this section, we obtain approximate analytic expressions for $\bar f$ and $C_{\d f(t)}$ in terms of MP number and properties. 

\subsubsection{Mean force}
\label{sec_mean_force}
Using mean-field approximation, considering each MP to be identical and independent, we first obtain an estimate of $\bar f$.  
The dynamical equations can be written as~(Appendix-\ref{app_mft})
\bea
\la \dot n_a \ra &=& (N-\la n_a \ra) \on - \la n_a \ra \w_d e^{k_m  \la y \ra/f_d}, \nn\\
 \la \dot y \ra &\approx& v_0 (1 - k_m \la y \ra/f_s) +  \la \dot x \ra, \nn\\
\g_f  \la \dot x \ra &=& - \la n_a \ra k_m \la y \ra + F_e,
\label{eq_MFT_EOM}
\eea
where $y = (1/n_a)\sum_{i=1}^{n_a} y_i$ denotes the arithmetic mean of MP extensions. 
We further replaced the mean detachment rate $\w_0=\w_d \la e^{k_m  y /f_d} \ra$ with the lower bound $\w_d e^{k_m  \la y \ra /f_d} $~(Appendix-\ref{app_mft}).

At the stall condition $\la \dot y \ra=0$, Eq.(\ref{eq_MFT_EOM}) gives
\bea
\g_f \la \dot x \ra = - \f{\la n_a \ra f_s}{1 + \f{\la n_a \ra f_s}{\g_f v_0}} + \f{F_e}{1 + \f{\la n_a \ra f_s}{\g_f v_0}},
\label{eq_xdot_MFT}
\eea
a behavior similar to that in Ref.~\citenum{Klumpp2005}. 
Here the first term on the right-hand side is the mean active force $\la f \ra$ due to MPs. The scaled dimensionless form $\tilde f = \la f \ra/\g_f v_0$ can be expressed as, 
\bea
\tilde f = -\f{\la n_a \ra \tilde f_s}{1+ \la n_a \ra \tilde f_s}, 
\label{eq_fbar}
\eea
where $\tilde f_s = \f{f_s}{\g_f v_0}$. In the absence of external force, this relation gives the scaled mean velocity of the filament $v/v_0=\tilde v$. It increases with the number of associated MPs to saturate. 

The steady-state estimate of the number of attached MPs 
\bea
\bar n_a = \f{\on}{(\on+\w_0)} N = \o N
\label{eq_nas}
\eea
can be obtained by setting $\la \dot n_a \ra= 0$, where $\w_0 = \w_d \la e^{k_m y/f_d}\ra$ and $\o={\on}/{(\on+\w_0)}$ denotes the processivity. 
In Fig.\ref{fig_results1}($a$) we show the variation of $\tilde f$ with $N$ using the processivity $\o=0.52$ obtained for the unloaded filament~(see Fig.\ref{fig_na_y}($a$) in Appendix-\ref{app_1}). 
The qualitative feature of the numerical observation agrees with the analytic expression. 
However, simulation results show a smaller value than the analytic estimate. The saturation value of numerically obtained $\tilde f$ remains smaller than the prediction of saturation $-\tilde f=1$ obtained from Eq.(\ref{eq_fbar}). The reason for the discrepancy will be considered carefully in Sec.~\ref{sub_sec_self_load}. 

Despite the non-linearity of the mean filament velocity, it is remarkable that the stall force of the filament $F_e^{s}=\la n_a \ra f_s$, obtained by using $\la \dot x \ra=0$ in Eq.(\ref{eq_xdot_MFT}), remains proportional to the number of MPs in agreement with earlier results~\cite{Klumpp2005, Kolomeisky2007a, Bhat2017}. Thus the study of the stall force of filament gives a good measure for the number of cargo-bound MPs~\cite{Soppina2009, Rai2013b}.

Moreover, Eq.(\ref{eq_xdot_MFT}) gives a prediction for the effective viscous drag 
in the presence of MPs
\bea
\g_{\rm eff} = \g_f (1+\la n_a \ra \tilde f_s). 
\label{eq_geff}
\eea
Similar linear growth in the drag coefficient was observed in phenomenological models studied earlier~\cite{Leibler1993, Nakul2021}.
Note that the linear growth of $\g_{\rm eff}$ with the mean number of active MPs differ qualitatively from the exponential
growth due to passive cross-linkers~\cite{Lansky2015}.

Further, comparing Eq.(\ref{eq_MFT_EOM}) with Eq.(\ref{eq_xdot_MFT}) we get an estimate of the mean  extension at stall 
\bea
k_m \bar y = f_s \f{1+\tilde f_e}{1+\la n_a \ra \tilde f_s}
\label{eq_yavg}
\eea
where $\tilde f_e=\f{F_e}{\g_f v_0}$. This expression 
shows good agreement with the simulation results~(see Fig.\ref{fig_na_y}($b$) in Appendix-\ref{app_1}). 

\subsubsection{Force correlation}
\label{sec_force_corr}

The two-time correlation of force fluctuations is given by $k_m^{-2} \la \d f(t) \d f (t') \ra =  \la \sum_{i,j} \d y_i (t) \d y_j (t')\ra$. 
Determination of a closed-form analytic expression for this correlation function is challenging. 
In the presence of precise synchrony between the extensions of different MPs, one can replace 
$\d y_j(t) = \d y_i(t)$~\footnote{In the absence of synchrony $\la \d y_i (t) \d y_j (t')\ra= \d_{ij} \la \d y_i (t) \d y_i (t') \ra$. We discuss later the impact of reduced synchrony in a self-load generation.}. 
This leads to $k_m^{-2} \la \d f(t) \d f (t') \ra = \la \sum_{i,j} \d y_i (t) \d y_i (t') \ra$. To estimate the correlation between MP extensions, we proceed as follows.

Relaxation of MP length in the attached state can be analyzed by combining the second and third equations in Eq.(\ref{eq_MFT_EOM}), giving 
\bea
\la \dot y \ra = \left( v_0 + \f{F_e}{\g_f}\right) - \f{\la y \ra}{\t_{el}}
\label{eq_meanydot}
\eea
with an elastic relaxation time 
\bea
\t_{el} = \f{1}{\tilde k_m} \left( \f{\tilde f_s}{1+\la n_a \ra \tilde f_s}\right) \w_d^{-1}
\label{eq_tau_el}
\eea 
where $\tilde k_m=k_m/\g_f \w_d$. 

The stochastic motion of each unloaded MP in the attached state can be treated as a Poisson process in which the MP moves in a directed fashion with a stochastic rate $\a=v_0/\s$ where $\s$ is the MP step-size. The probability $P_m$ for the MP to be at $m$-th site at time $t$ evolves as  $dP_m/dt= \a P_{m-1} - \a P_m$ with the initial condition $P_m(t=0)=\d_{m,0}$. The solution gives the Poisson distribution $P_m(t) = e^{-\a t} (\a t)^m/m!$.  Thus the fluctuation in displacement $\la \d y^2 (t)\ra = \s^2 \la\d m^2\ra$ where $\la\d m^2\ra = [\la m^2 \ra - \la m \ra^2] = \a t$, as $\la m\ra=\a t$, $\la m^2\ra=(\a t)^2 + \a t$. Writing $\la \d y^2 \ra = 2  D_y t$, we get the expression for effective diffusivity for each MP  around the mean drift
\bea
D_y  = v_0 \s/2.
\label{ep_mp_diff}
\eea
To obtain an estimate of the correlation in the arithmetic mean extension $y$ of $\la n_a \ra$ flexible linkers corresponding to the MPs, we add the stochasticity mentioned above,  arising from the Poisson process to the mean-field dynamics Eq.(\ref{eq_meanydot}). In the absence of external force, this leads to the following Ornstein-Uhlenbeck process
\bea
\dot y = v_0 - y/\t_{el} + (2D_y/\la n_a \ra)^{1/2}\, \eta(t)
\label{lange_y}
\eea
where the white noise obeys $\la \eta(t)\ra=0 $, $\la \eta(t) \eta(t')\ra = \d(t-t')$. Considering the extensions of MPs as independent random variables, the standard deviation of their sum grows as $\sqrt{\la n_a \ra}$. This led to the $\la n_a \ra^{-1/2}$ decay in the fluctuation strength $(2D_y/\la n_a \ra)^{1/2}$ around the mean extension. The mean-field limit is obtained for large $N$. 
It is straightforward to solve the Langevin equation Eq.(\ref{lange_y}) to find
\bea
\la y(t) y(t') \ra = \la y \ra^2 + (D_y \t_{el} /\la n_a \ra) e^{-|t-t'|/\t_{el}} ,
\label{eq_ytyt_corr_full}
\eea 
which leads to
\bea
\la \d y(t) \d y(t') \ra = (D_y \t_{el} /\la n_a \ra) e^{-|t-t'|/\t_{el}} ,
\label{eq_ytyt_corr}
\eea  
for a given $\la n_a \ra$. 
We proceed by replacing the extension correlation for $i$-th MP $\la \d y_i (t) \d y_i (t') \ra$ with the correlation for the arithmetic mean extension $\la \d y(t) \d y(t') \ra$. 

Within this approximation, $\la \sum_{i,j} \d y_i (t) \d y_i (t') \ra \approx \la n_a(t) \ra^2 \la \d y(t) \d y(t') \ra$ using the above expression for 
extension correlation obtained for a given $\la n_a \ra$. 
The time scale $\t_{ad}$ determining the correlation $\la n_a(t) n_a(t')\ra$~(see Appendix-\ref{app_corr}) is much longer than the relaxation time $\t_{el}$, allowing the above approximation in which $\la n_a\ra$ is held fixed. Within this approximation, the force and its fluctuations arise essentially due to the extension of MPs attached to the filament. 
The correlation in force fluctuation can then be expressed as
\bea
C_{\d f(t)}=\la \d f(t) \d f(0)\ra \approx C e^{-t/\t_{el}}
\label{eq_dftdft}
\eea
with
\bea
C = \la{n_a} \ra k_m^2 D_y \t_{el},
\label{eq_C0}
\eea
where the steady-state estimate for the number of attached MPs can be used instead of $\la n_a \ra$ for comparison with the numerical results. 

In deriving this expression, we used the fluctuation of extension of attached MPs, neglecting the relatively slow evolution of the attachment detachment process. At attachment, MPs remain unextended; as a result, it does not change the force immediately. However, the detachment of extended MPs can cause significant force fluctuations. We could not incorporate this mechanism within our simple analytic approach. Our estimated strength of force fluctuation $C$ given by Eq.(\ref{eq_C0}) remains smaller than the numerical observations~(Fig.\ref{fig_results1}($d$)\,). It is possible to use Fokker-Planck equations describing MPs' attachment- detachment- extension dynamics and calculate the correlation functions directly. However, getting a complete closed-form analytic expression remains challenging.  

\begin{figure}[t]
\includegraphics[width=8.6cm]{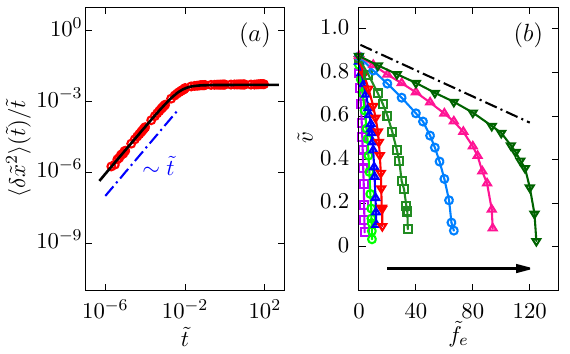}
\caption{ (color online) Displacement fluctuation and force-response. We use dimensionless quantities $\tilde{x} = x/(v_0 \w_d^{-1})$,  $\tilde{t} = t \w_d$, $\tilde{v} = v/v_0$ and $\tilde{f}_e = F_e/(\g_f v_0)$. $(a)$~Points denote simulation results and the black solid line plots Eq.(\ref{eq_mean_dx2t_effective}) for a filament driven by $N=8$ MPs. While plotting we use $\la n_a \ra = \bar n_a$ in the expression of $\t_{el}$. 
$(b)$ Velocity response of force for the filament driven by $N = 4,8,16,40,80,120,160$ MPs with the arrow denoting the direction of increasing $N$.  Velocity decreases linearly for small forces (e.g., the dashed line) before the onset of non-linear decrease at a higher load.}
\label{fig_Deff_mueff}
\end{figure}
 
The above expression captures the exponential decay of correlation functions in Fig.(\ref{fig_results1})($b$). 
The correlation
time $\t=\t_{el}$ decreases with $N$ following Eq.(\ref{eq_tau_el}). This estimate shows excellent 
agreement with simulation results for correlation time shown in Fig.(\ref{fig_results1})($c$). 
According to the above estimate, the dimensionless force fluctuation 
\bea
\tilde C = \f{C}{(\g_f v_0)^2} = \f{\s k_m}{2 \g_f v_0}  \f{\la n_a\ra \tilde f_s}{1+\la n_a \ra \tilde f_s}
\label{eq_C}
\eea
grows with $N$ to saturate. This feature agrees qualitatively with the simulation results presented in Fig.(\ref{fig_results1})($d$).
We verified from numerical simulations that $C \propto v_0$.

The calculation of the mean force in Eq.(\ref{eq_fbar}), force fluctuation strength $\tilde C$ in Eq.(\ref{eq_C}) and the correlation time $\t=\t_{el}$ in Eq.(\ref{eq_tau_el}) completes the description of MP driven filament motion as an active Ornstein-Uhlenbeck process given by Eq.(\ref{eq_xdot_effective}). This is the first main achievement of this paper.

\subsection{Fluctuation, response, and effective temperature}
\label{sub_sec_FDT}
The Langevin equation Eq.(\ref{eq_xdot_effective}) can be directly solved to show that the mean displacement increases with time as $\la x \ra = \f{\bar f}{\g_f} t$ and the mean-squared deviation $\la \d x^2\ra = \la x^2\ra - \la x \ra^2$ shows
\bea
\la \d x^2 \ra (t) 
= \f{2 C \t_{el}}{\g_f^2} \left[ t - \t_{el} (1 - e^{-t/\t_{el}}) \right]
\label{eq_mean_dx2t_effective}
\eea
using the correlation time $\t=\t_{el}$. 

\begin{figure}[t]
\begin{center}
\includegraphics[width=8.6cm]{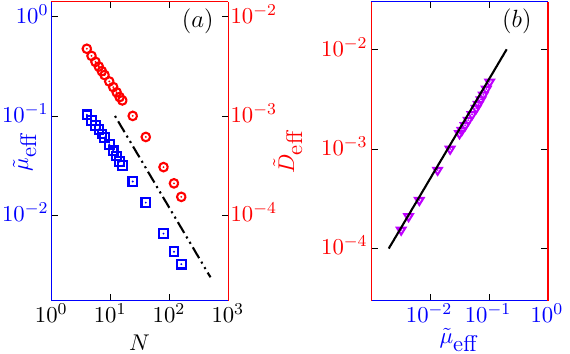}
\caption{ (color online) Mobility, diffusivity and effective temperature. We plot dimensionless quantities $\tilde{\mu}_{\rm eff} = \mu_{\rm eff}\g_f$ and $\tilde{D}_{\rm eff} = D_{\rm eff}/(v_0^2 \w_d^{-1})$. 
$(a)$~Points ${\color{blue} \boxdot}$ (${\color{red} \odot}$) denote simulation results for mobility (asymptotic diffusivity). The dashed line depicts a $N^{-1}$ scaling. 
$(b)$~Points denote numerical results for $\tilde{D}_{\rm eff}$ and $\tilde{\mu}_{\rm eff}$. The black solid line plots $D_{\rm eff} = \mu_{\rm eff} \,T^{\infty}_{\rm eff}$.}
\label{fig_FDT}
\end{center}
\end{figure}

This expression predicts a ballistic-diffusive cross-over around $t=\t_{el}$ such that,
$\la \d x^2 \ra \approx \f{C}{\g_f^2} t^{2} $ for $t \ll \t_{el}$ and in the long-time limit of $t \gg \t_{el}$
$\la \d x^2 \ra \approx 2 D_{\rm eff} t $~(see Fig.\ref{fig_Deff_mueff}($a$)) with 
\bea
D_{\rm eff} = \f{C \t_{el}}{\g_f^2}=\f{\la n_a\ra \tilde f_s^2}{(1+ \la n_a \ra \tilde f_s)^2}\, \f{\s v_0}{2}.
\label{eq_Deff}
\eea 
Here, the attached MPs lead to displacement fluctuations. 
In the absence of an explicit translational noise, $D_{\rm eff}=0$ when $N=0$. On the other hand, for large $N$, the effective diffusion constant decreases as $N^{-1}$. Such large $N$ dependence of the effective filament diffusivity agrees with the earlier estimate in Ref.~\citenum{Leighton2022}. Direct numerical simulation results presented in Fig.\ref{fig_FDT}($a$) concur with this prediction. 

The reliability in stochastic transport, in the presence of such fluctuations,  can be quantified by the asymptotic coefficient of variation $\h = \sqrt{\la \d x^2 \ra}/\la x \ra$~\cite{Leighton2022, Mugnai2020} or the  Fano factor $\phi = \la \d x^2 \ra/\la x \ra$~\cite{Brown2019}. We find
\bea
\h = \left( \f{\s}{\la n_a \ra v_0}\right)^{1/2} t^{-1/2} ,
\eea
showing that $\h$ decreases with both the chemical activity $v_0$ and the number of MPs, increasing transport reliability.  The asymptotic displacement Fano factor measuring fluctuations in transport
\bea
\phi = \f{\s \tilde f_s}{1+ \la n_a \ra \tilde f_s}
\eea
reduces as $N^{-1}$ for a large number of MPs.

The velocity response of the filament due to external load force acting against the MP drive is obtained from numerical simulations and shown in Fig.\ref{fig_Deff_mueff}($b$). The filament velocity decreases with the load, first linearly and then more sharply at a larger load. The figure shows that the force response strongly depends on the number of MPs acting on the load. The mobility at zero load can be obtained numerically from the slope of the force-velocity graph Fig.\ref{fig_Deff_mueff}($b$) near $\tilde f_e=0$. The analytical estimate of the effective mobility $\mu_{\rm eff}=\g_{\rm eff}^{-1}$ 
is given by Eq.(\ref{eq_geff}) and has the form
\bea
\mu_{\rm eff} 
= \g_f^{-1}\f{1}{1 + \la n_a \ra \tilde f_s}.
\label{eq_mueff}
\eea 
This predicts that for large $N$ the mobility should scale as $N^{-1}$, in agreement with the numerical simulation result for $\mu_{\rm eff}$ shown in   Fig.\ref{fig_FDT}($a$).  

The equilibrium Einstein relation connecting mobility and diffusivity via bath temperature does not generally hold in systems out of equilibrium. Even in a steady state, the generalized fluctuation-response relation involves an additive correction that depends on the steady-state current~\cite{Agarwal1972, Seifert2010, Cugliandolo2011, Chaudhuri2012, Geiss2020}. 
As we found from numerical simulations, the diffusivity of the filament remains proportional to its mobility at all values of activity. This allows us to
use the ratio of long-time diffusivity and mobility, Eq. (\ref{eq_Deff}) and (\ref{eq_mueff}), to define an effective temperature,
$T_{\rm eff} = \f{D_{\rm eff}}{\mu_{\rm eff}}$ which can be expressed as
\bea
\f{T_{\rm eff}}{T_{\rm eff}^\infty} = \f{\la n_a \ra \tilde f_s}{1+\la n_a \ra \tilde f_s},~
{\rm where,}~ T_{\rm eff}^\infty = \f{\s f_s}{2} 
\eea
is the $N$-independent, effective temperature obtained in the large $N$ limit. This is given by the energy dissipation $\f{\s f_s}{2}$ by MPs per motor cycle. 
The line in Fig.~\ref{fig_FDT}($b$) shows that $D_{\rm eff}$ approaches $\mu_{\rm eff} T_{\rm eff}^\infty$ asymptotically for large $N$.  In this case, the short correlation time $\t_{el}$ in fluctuations of active force allows for an estimate of equilibrium-like effective temperature,
%
which will be utilized in the following section to describe the self-load.

\begin{figure}[t] 
\begin{center}
\includegraphics[width=8.6cm]{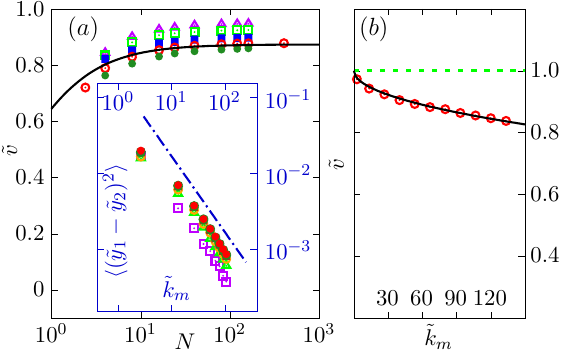} 
\caption{ (color online) Dependence of filament velocity on MP number and stiffness. We use dimensionless quantities $\tilde{k}_m = k_m/(\g_f \w_d)$, $\tilde{y}_i = y_i/(v_0 \w_d^{-1})$, and $\tilde{v} = v/v_0$. 
$(a)$  $\tilde{v}$ grows and saturates with $N$. Different point types denote different values of 
$\tilde k_m=13.33$\,({\color{darkmagenta} $\triangle$}), 
$26.67$\,({\color{green} $\boxdot$}),
$53.33$\,({\color{blue} $\blacksquare$}),
 $80$\,({\color{red} $\odot$}),
 $106.67$\,({\color{forestgreen} \LARGE{$\bullet$}}). 
The black solid line plots Eq.(\ref{eq_v_vs_N_with_fsl}) using Eq.(\ref{eq_fsl}) with $\a=0.95$.
In this plot we use $\la n_a \ra = \bar n_a$.
Inset: Relative fluctuations in extension for  
$N = 4$\,({\color{darkmagenta} $\boxdot$}) 
$8$\,({\color{green} $\triangle$})
$16$\,({\color{orange} $\odot$})
$160$\,({\color{red} \LARGE{$\bullet$}})
The blue dash-dotted line shows a $1/k_m$ scaling.
$(b)$~Points denote simulation results for $\tilde{v}$ as a function of $\tilde{k}_m$ for $N=160$. The black solid line plots  
$\tilde v = 1-f_{sl}/f_s$ where the asymptotic value of self-load $f_{sl}=\sqrt{2(1-\a)k_mT_{\rm eff}^\infty}$ with $\a=0.95$. 
}
\label{fig_result2}
\end{center}
\end{figure}

\subsection{Cooperativity and self-load}
\label{sub_sec_self_load}
Finally, we return to the dependence of filament velocity on the number of MPs. This is shown for MPs with different spring constants $k_m$ in Fig.~\ref{fig_result2}($a$). The saturation values remain smaller than the mean-field prediction of $-\tilde f/\g_f$ with $\tilde f$ given by Eq.(\ref{eq_fbar}) and reduces further with increasing $k_m$. This behavior is due to the generation of a self-load resulting from 
a lack of synchrony
between different MP extensions.

At this stage, let us assume a local thermodynamic equilibrium to use the effective temperature $T_{\rm eff}$, characterizing the active filament fluctuations, to determine the amount of fluctuation in $i$-th MP extension $\la y_i^2\ra = T_{\rm eff}/k_m$. 
This approximation shows reasonable agreement with the numerical evaluation of $\la y_i^2\ra$ as a function of $k_m$~(See Fig.\ref{fig_na_y} in Appendix-\ref{app_1}). 
The relative extension of MPs can be expressed as $\la (y_1-y_2)^2\ra = 2(1-\a) \la y_1^2\ra$ using 
$\la y_1 y_2\ra=\a \la y_1^2\ra$ where $\a$ quantifies the amount of synchrony between extensions of two MPs. This leads to 
\bea
\la (y_1 - y_2)^2 \ra = \f{2(1-\a) T_{\rm eff}}{k_m}.
\label{eq_Teff_equipartition}
\eea
If the extensions are perfectly in synchrony, $\a=1$, which gives $\la (y_1 - y_2)^2 \ra =0$. However, in general $\a<1$.  
The inset of Fig.~\ref{fig_result2}($a$) shows that relative fluctuations obtained from numerical simulations indeed varies as $\la (y_1 - y_2)^2 \ra \sim 1/k_m$.
The self-load due to the relative fluctuation has a measure $f_{sl} = k_m y_{sl}$ with $y_{sl} = \la (y_1 - y_2)^2 \ra^{1/2}$. As a result, we obtain $f_{sl} = [2(1-\a)k_m\, T_{\rm eff}]^{1/2}$, which can be expressed as
\bea
f_{sl} = \sqrt{2 (1-\a) k_m T_{\rm eff}^\infty} \left( \f{\la n_a \ra \tilde f_s}{1+\la n_a \ra \tilde f_s}\right)^{1/2}.
\label{eq_fsl}  
\eea 
In Eq.(\ref{eq_xdot_MFT}), replacing external load $F_e$ by the net self load $\la n_a \ra f_{sl}$ due to $\la n_a \ra$ MPs,  we obtain the following relation for the filament velocity
\bea
\tilde v = -\f{\dot x}{v_0} = \f{\la n_a \ra (\tilde f_s-\tilde f_{sl})}{1+\la n_a \ra \tilde f_s} ,
\label{eq_v_vs_N_with_fsl}
\eea
where $\tilde f_{sl} = f_{sl}/\g_f v_0$. In the absence of self-load, the active force acting on the filament arises due to the MP stall force $f_s$. The self-load $f_{sl}$ acts against this active force to reduce its impact. 

Using $\a=0.95$, the expression in Eq.(\ref{eq_v_vs_N_with_fsl}) along with Eq.(\ref{eq_fsl}) captures the dependence of filament velocity $\tilde v$ on number of MPs $N$, for all values of $k_m$~(see Fig.~\ref{fig_result2}($a$)\,). Thus the quantitative loss in synchrony is $5\,\%$ and is independent of $k_m$. 
Moreover, Eq.(\ref{eq_fsl}) shows that $f_{sl} \sim k_m^{1/2}$. This expression captures the decrease in $\tilde v$  with increasing $k_m$, as shown in Fig.~\ref{fig_result2}($b$). In the limit of large $N$, the expression simplifies to $\tilde v = 1-f_{sl}/f_s$. The solid line in Fig.~\ref{fig_result2}($b$) plots this expression with $\a=0.95$.  
Similar reductions in velocity with increasing spring stiffness have been recently observed in numerical simulations presented in Ref.[\citenum{Blackwell2019}]. Our theory provides a clear explanation of such observations.
The deviation of the saturation value of $\tilde v$ from unity is given in terms of $\sqrt{(1-\a)\f{k_m\s}{f_s}}$. Apart from the lack of synchrony $\a$, it is controlled by the spring constant $k_m$, step-size $\s$, and the stall force $f_s$. 

The identification and estimation of the self-load, 
and the determination of its impact on filament velocity is the second main result of this paper.
\section{Discussion}
\label{sec_discussion}

In this paper, we developed an active bath picture to describe the collective impact of motor proteins (MP) on a conjugate filament. This provides an effective Langevin dynamics with active mean force and force fluctuation that can be described as an active Ornstein-Uhlenbeck process. 
We derived the expression of the mean force using a mean-field analysis. Approximate analytic expressions of the force fluctuation amplitude and the force correlation time are also obtained. As we have shown, the force fluctuations are essentially governed by the fluctuations of MP extension in the attached state.     

Solving the effective Langevin equation describing the motion of filament under MP drive, we found the asymptotic diffusivity and mobility of the filament. This led to an effective temperature that grew and saturated with the MP number. The stall force and MP step size entirely determine the saturation.  Within  a local thermal equilibrium approximation, the effective temperature also describes the fluctuation of MP extension. Using this, we estimated the relative fluctuations of MP extensions which would have vanished if the individual extensions were in perfect synchrony. In the absence of that, an effective self-load emerges. 
We obtained an approximate expression of this self-load that describes well why the filament velocity under the drive of a large number of MPs saturates to a value smaller than that of a free MP.     

The main assumptions involved in the approximate analytic calculations are (i)~The active force on the filament is due to attached MPs extending until reaching the stall condition. Thus the mean active force is obtained by using the stall condition $\langle \dot y \rangle=0$.  (ii)~The active force fluctuations are essentially determined by MP extensions. (iii)~An equilibrium-like effective temperature is obtained from the ratio of diffusivity and mobility. This is used in a local thermal equilibrium argument to determine self-load expression. At attachment, the MPs do not extend, and as a result attachment process can produce slight force fluctuation. In contrast, extended MPs can cause significant force fluctuation at the detachment. This effect could not be incorporated in our approximate analytic expression for force correlation. This led to a smaller estimate of force fluctuation with respect to numerical observations.

The parameter values used in our numerical simulations correspond to the microtubule-kinesin systems. 
Thus our quantitative predictions are amenable to direct experimental measurements in such systems. 
However, the scheme presented here is generic and is equally applicable to other MP-filament systems, e.g., actin filament-myosins~\cite{Kron1986, Harada1990, Toyoshima1990}. 
Developing an effective active bath picture for the filament motion in MP assay is also relevant to the recent interest in tracer dynamics in active baths~\cite{Granek2021a, Ye2020}.  
As has been shown before, in experiments, the number of MPs can be precisely controlled~\cite{Derr2012, Furuta2013}, thus allowing for testing the MP-number dependences predicted in our study. For example, in Ref.[\citenum{Lam2014}]  kinesin surface density was varied between $90 \pm 40$ $\mu$m$^{-2}$ and $1600 \pm 580$ $\mu$m$^{-2}$.
This range of surface density corresponds to a line density 
range of MPs from $9.5 \pm 6.3\,\mu$m$^{-1}$ to $40 \pm 24\, \mu$m$^{-1}$. 
The line density of MPs used in such experiments can be varied in the range of $3$ $\mu$m$^{-1}$~\cite{VanDelinder2019}
to $120$ $\mu$m$^{-1}$~\cite{VanDelinder2016}.
Note that we varied the line density in our numerical study between $0.5$ to $50\,\mu$m$^{-1}$.  The amount of self-load generated in experiments can be determined from the knowledge of the unloaded self-propulsion of MPs. 

Moreover, our method can be utilized in coarse-grained theoretical studies of semiflexible bio-polymers in gliding assays~\cite{Shee2021, Gupta2019a}. The active bath mapping developed here can simplify numerical calculations by removing the requirement of simulating all the MPs and treating the filament as locally active driven by colored noise whose properties are determined by the MPs. This can allow a more detailed investigation of the impact of MP activity on bio-polymer assemblies.

\section*{Author Contributions}
DC designed the study. CK performed numerical and analytical calculations under the supervision of DC. DC and CK wrote the paper.

\section*{Conflicts of interest}
There are no conflicts to declare.

\section*{Acknowledgements}
We thank Dibyendu Das of IIT-Bombay and Debashish Chowdhury of IIT-Kanpur for valuable discussions. D.C. thanks SERB, India, for financial support through Grant No. MTR/2019/000750 and International Centre for Theoretical Sciences (ICTS) for an associateship. The numerical simulations were partly supported by SAMKHYA, the High-Performance Computing Facility provided by the Institute of Physics, Bhubaneswar.

\appendix
\section{Derivation of mean field equations}
\label{app_mft}
Defining the arithmetic mean of the individual MP extensions as $y = (1/n_a) \sum_i ^{n_a} y^i $ one obtains the evolution of the mean number of attached MPs as~\cite{Shee2021a}
\bea
\la \dot n_a \ra &=& (N-\la n_a \ra) \on - \la n_a \w_d  e^{k_m y/f_d} \ra,
\eea
where $\la \dots \ra$ denotes statistical averaging over stochastic processes. Within mean-field approximation, in the above equation, we first replace $\la n_a e^{k_m y/f_d} \ra = \la n_a \ra \la e^{k_m y/f_d} \ra$. By Jensen's inequality, $\la e^{k_m y/f_d} \ra  \geq e^{k_m \la y \ra/f_d}$. Thus the real relaxation of $\la n_a \ra$ is faster than that assumed in Eq.(\ref{eq_MFT_EOM}). 

Using Eq.(\ref{eq_ydot_ith}) within the linear regime of force-velocity, one gets the second equation in  Eq.(\ref{eq_MFT_EOM}). 
Further, setting the external force $F_e=0$, writing $\la F_m \ra = -k_m \la \sum_{i=1}^{n_a} y^i \ra = -\la n_a \ra k_m \la y \ra$ within the mean-field approximation in Eq.(\ref{eq_xdot_force_balance}), one obtains  the third equation in Eq.(\ref{eq_MFT_EOM}).

\begin{figure}[t] 
\includegraphics[width=8.6cm]{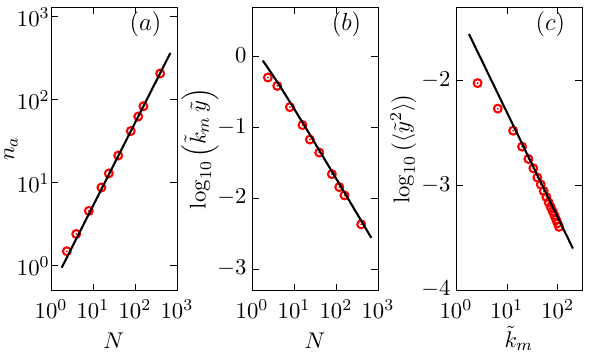}
\caption{ (color online) $(a)$~Points denote simulation results for $\bar n_a$ as a function of $N$. The black solid line plots $\bar n_a = \o N$ with the processivity $\o=0.52$. 
$(b)$~The scaled extension $\tilde{k}_m\,\tilde{y} = k_m\,y/(\g_f v_0)$ obtained from numerical simulations are shown by points. The black solid lines plots Eq.(\ref{eq_yavg}). While plotting we use $\la n_a \ra = \bar n_a$.
$(c)$ Points denote simulation results for scaled mean squared extension $\la \tilde{y}^2 \ra = \la y^2 \ra/(v_0 \w_d^{-1})^2$ as a function of scaled motor stiffness $\tilde{k}_m = k_m/(\g_f \w_d)$, for $N=40$. The black solid line plots  $\la \tilde y^2 \ra = \f{\tilde T_{\rm eff}^{\infty}}{\tilde k_m}$ where $\tilde T_{\rm eff}^{\infty}=T_{\rm eff}^{\infty}/(\g_f v_0^2 \w_d^{-1})$.}
\label{fig_na_y}
\end{figure}

\section{Processivity and extension}
\label{app_1}
From numerical simulations using parameter values listed in Table-\ref{tab_parameter_values} we calculate the steady-state mean number of attached MPs $\bar n_a$. It grows linearly with the total number of MPs $N$ with a slope $\o$ giving the processivity in the absence of external load~(Fig.\ref{fig_na_y}($a$)). In Fig.\ref{fig_na_y}($b$) we plot the mean extension of MPs that decreases with $N$ showing excellent agreement with the prediction in Eq.(\ref{eq_yavg}). Further, in Fig.\ref{fig_na_y}($c$) we show the simulation results for the mean-squared extension of MPs and compare them with the approximate estimate $\la y^2 \ra = T_{\rm eff}/k_m$ presented in Sec.~\ref{sub_sec_self_load}.      

\begin{figure}[t] 
\includegraphics[width=8.6cm]{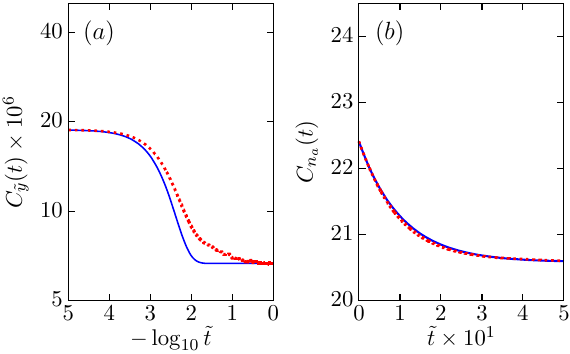}
\caption{(color online) Correlation functions. We use dimensionless quantities $\tilde{t} = t \w_d$, $C_{\tilde{y}}(t) = \la y(t) y(0) \ra (\w_d^2/v_0^2)$, $C_{n_a}(t) = \la n_a(t) n_a(0) \ra$. 
$(a)$~The red dashed line plots $C_{\tilde{y}}(t)$ obtained from numerical simulations.
The blue solid line plots $C_{\tilde{y}}(t)$ using Eq.(\ref{eq_ytyt_corr_full}) with $\t_{el}$ from Eq.(\ref{eq_tau_el}) with $\la n_a \ra = \bar n_a$ and mean and standard deviation from numerical results. 
 $(b)$~The red dashed line plots $C_{n_a}(t)$ obtained from  from numerical simulation. The solid blue line plots the same quantity using the expression in  Eq.(\ref{eq_nat_nat_corr}) with $\t_{ad} = 1/(\on + \w_0)$ and mean and standard deviations from numerical results.}
\label{fig_corr}
\end{figure}

\section{Correlation functions}
\label{app_corr}

In the attempt to obtain a closed-form expression for force correlation, we focussed on the fluctuation of  $y(t) = \f{1}{n_a(t)} \sum_{i=1}^{n_a(t)} y_i(t)$ for a fixed $\la n_a \ra$. The expression in Eq.(\ref{eq_dftdft}) captures the correlation time and qualitative features of the strength of fluctuations $\tilde C$ as shown in Fig.\ref{fig_results1}. The expression of $\la y(t) y(t')\ra$ in Eq.(\ref{eq_ytyt_corr}) shows semi-quantitative agreement with simulation results~(Fig.\ref{fig_corr}($a$)). In Fig.\ref{fig_corr}($b$) we plot the auto-correlation of number of attached MPs.

The attachment-detachment can be considered a random Telegraph process with rates $\on$ and $\w_0$, where  $\w_0=\w_d \la e^{k_m  y /f_d} \ra$ is calculated from the direct numerical measurements of $\la n_a\ra = N P^s_a$, where the steady-state probability of attached fraction $P^s_a=\on/(\on + \w_0)$. The variance is given by $\d n_a^2= \la n_a^2\ra - \la n_a\ra^2=\f{\on \w_0}{(\on + \w_0)^2} N$. The Telegraph process predicts a steady-state correlation~\cite{Gardiner_book}
\bea
\la n_a(t) n_a(t') \ra = \la n_a\ra^2 + \d n_a^2\, e^{-|t-t'|/\t_{ad}}.
\label{eq_nat_nat_corr}
\eea
where $\t_{ad} = 1/(\on + \w_0)$.
The simulation results in Fig.\ref{fig_corr}($b$) show excellent agreement with the analytical prediction in Eq.(\ref{eq_nat_nat_corr}).

\bibliographystyle{rsc}

\providecommand*{\mcitethebibliography}{\thebibliography}
\csname @ifundefined\endcsname{endmcitethebibliography}
{\let\endmcitethebibliography\endthebibliography}{}

\end{document}